\documentclass[aps,prb,superscriptaddress,twocolumn]{revtex4-2}
\usepackage{graphics}
\usepackage{epsfig}
\usepackage{color}
\usepackage{stackrel}
\usepackage{amsfonts}
\usepackage{wrapfig}
\usepackage{pstricks}
\usepackage{pst-node}
\usepackage{bm}
\usepackage{dcolumn}
\usepackage{multirow}
\usepackage{epstopdf}
\usepackage{subfigure}
\usepackage{amssymb}
\usepackage{amsmath}
\usepackage{commath}
\usepackage{graphicx,bm}
\usepackage{verbatim}
\usepackage{booktabs}
\usepackage[para]{threeparttable}
\usepackage{etoolbox}
\usepackage{lipsum}
\usepackage{CJKutf8}
\usepackage{wasysym}
\usepackage{bbold}
\usepackage{xcolor}
\usepackage{hyperref}

\begin{document}

\title{Intervalley electron-hole exchange interaction and impurity-assisted recombination of indirect excitons in WS$_2$ and WSe$_2$ monolayers}

\author{Pengke Li (\begin{CJK*}{UTF8}{gbsn}李鹏科\end{CJK*})}
\altaffiliation{lipengke2021@gmail.com}
\affiliation{Department of Electrical and Computer Engineering, University of Rochester, Rochester, New York 14627, USA}

\author{Cedric Robert}
\affiliation{Universit\'{e} de Toulouse, INSA-CNRS-UPS, LPCNO, 135 Av. Rangueil, 31077 Toulouse, France}
\author{Dinh~Van~Tuan}
\affiliation{Department of Electrical and Computer Engineering, University of Rochester, Rochester, New York 14627, USA}
\author{Lei Ren}
\affiliation{Universit\'{e} de Toulouse, INSA-CNRS-UPS, LPCNO, 135 Av. Rangueil, 31077 Toulouse, France}
\author{Xavier Marie}
\affiliation{Universit\'{e} de Toulouse, INSA-CNRS-UPS, LPCNO, 135 Av. Rangueil, 31077 Toulouse, France}
\author{Hanan~Dery}
\altaffiliation{hanan.dery@rochester.edu}
\affiliation{Department of Electrical and Computer Engineering, University of Rochester, Rochester, New York 14627, USA}
\affiliation{Department of Physics and Astronomy, University of Rochester, Rochester, New York 14627, USA}

\begin{abstract}

The variety of excitonic states in  tungsten-based dichalcogenide monolayers stems from unique interplay between the spin and valley degrees of freedom. One of the exciton species is the  indirect exciton (momentum or valley dark), which is responsible to a series of resonances when the monolayer is charge neutral. We investigate the short-range electron-hole exchange interaction of the indirect exciton, as well as its recombination mechanism mediated by impurities. The analysis  provides thorough understanding of the energy and polarization of the zero-phonon indirect exciton resonance in the emission spectrum.
\end{abstract}

\maketitle  
\section{introduction}

\begin{figure}[t]
\includegraphics[width=8.4cm]{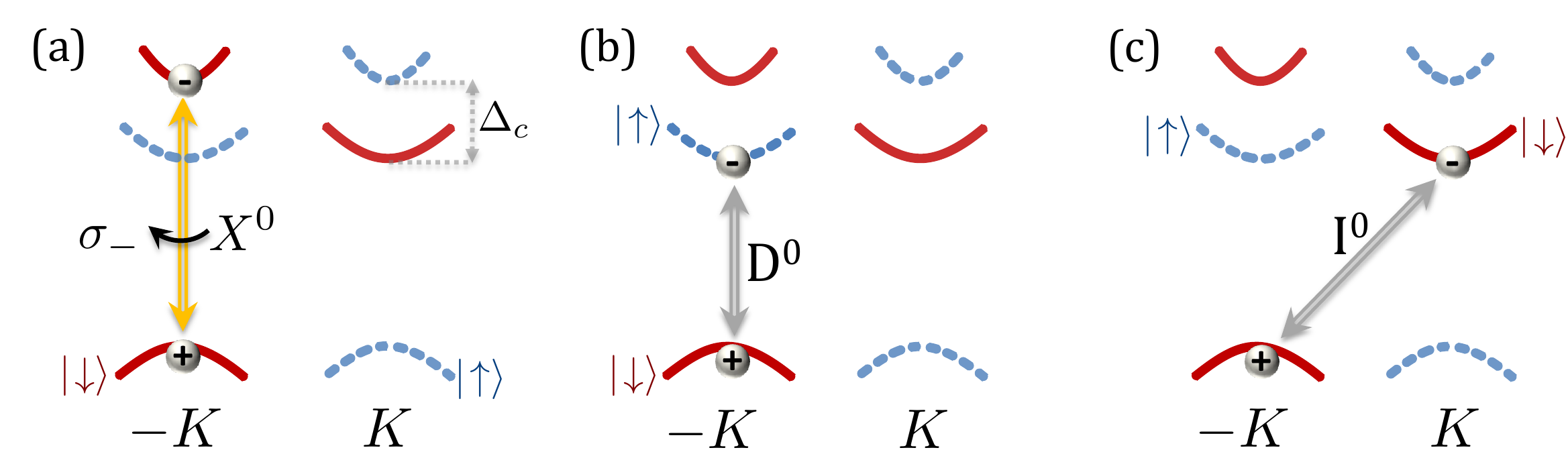}
\caption{Charge-neutral excitons in WSe$_2$ and WS$_2$ monolayers involving electronic states in valleys around the $\pm K$ point. Solid (dotted) lines correspond to energy bands with spin-down (spin-up) electronic states. The hole in the shown excitons resides in the top valley of the valence band at the $-K$ point. The resulting states are the bright exciton in (a), dark exciton in (b), and indirect exciton in (c). 
\label{fig:scheme}}
\end{figure}

Photoluminescence measurements of high quality WSe$_2$ reveal an exceptionally rich emission spectrum \cite{Robert_PRB17,Zhang_NatNano17,Barbone_NatComm18,Chen_NatComm18,Li_NatComm18,Ye_NatComm18,Liu_PRL19,Liu_PRL20,Tang_NatComm19,Li_NatComm19,Li_ACS19, He_NatComm20,Robert_PRL21}. Electrons of excitonic complexes in WSe$_2$ or WS$_2$ monolayers come with various spin and valley configurations, wherein optically active excitons (bright) reside in higher energies than excitons with weak optical transition \cite{Song_PRL13}. Optical excitation with circularly polarized light introduces valley-selective bright excitons \cite{Xiao_PRL12}, as shown in Fig.~\ref{fig:scheme}(a) for exciton composed of electronic states in the $-K$ valley. These photoexcited bright excitons then relax to the ground-state dark exciton, shown in Fig.~\ref{fig:scheme}(b), following phonon emission or exchange scattering with free electrons \cite{Dery_PRB15,Yang_PRB22}. Radiative recombination of the dark exciton involves an out-of-plane dipole moment \cite{Slobodeniuk_2DMater16}, which results in unpolarized light with equal emission amplitudes from right and left handed helicity ($\sigma_{\pm}$) \cite{Robert_PRB17,He_NatComm20}. The valley from which an optical transition of a dark exciton takes place can be identified by the circular polarization degree of its phonon replica \cite{He_NatComm20,Robert_PRL21,Yang_PRB22}, or by applying a magnetic field to lift the valley degeneracy  \cite{Robert_PRB17,Zhang_NatNano17}.

Detection of the indirect exciton, shown in Fig.~\ref{fig:scheme}(c),  is more elusive than detections of dark or bright excitons. Its energy is about 10 meV higher than that of the  dark exciton and its optical transition is much weaker than that of the bright exciton due to the indirect band gap \cite{He_NatComm20,Yang_PRB22}. The result is that light emission is mostly governed by the dark and bright excitons. The weak dipole transition of dark excitons is offset by their relatively large density because most excitons end up dark. As for the bright excitons, their relatively small density is offset by their strong dipole transition. Indirect excitons, on the other hand, are neither populous like dark excitons nor emissive like bright ones. 

The lifetime of indirect excitons is governed by exchange scattering with resident electrons that turn these excitons dark  \cite{Yang_PRB22}.  For example, exchange scattering of the indirect exciton in Fig.~\ref{fig:scheme}(c) with a spin-up free electron from the $-K$ valley can create a dark exciton in the $-K$ valley, as the one in Fig.~\ref{fig:scheme}(b), leaving behind a spin-down free electron in the $K$ valley. Cold dark excitons cannot turn indirect through a reversible exchange scattering process due to their lower energy.  Exchange scattering does not involve a spin-flip of the electron but an electron switch in the exciton. As such, exchange scattering is more relevant than exciton-phonon scattering in setting the dynamics between indirect and dark excitons. The phonon-assisted process requires an intervalley spin-flip scattering of the electron or hole, which to a leading order is suppressed by time-reversal symmetry \cite{Song_PRL13}. One way to amplify the emission intensity of indirect excitons is through valley pumping with circularly polarized light \cite{Robert_NatComm21}. The valley pumping selectively depletes free electrons from the photoexcited valley: light with $\sigma_+$ ($\sigma_-$) helicity depletes electrons from $K$ ($-K$). As a result, indirect excitons are less prone to become dark through exchange scattering, leading to stronger emission from indirect excitons \cite{Yang_PRB22}.

The goal of this work is to address two questions regarding the elusive indirect exciton. The first one deals with its spectral position: Why does the indirect exciton emerge $\sim$10 meV above the dark exciton? The reason cannot come from different band-gap energies because the energy gaps are identical for the dark and indirect excitons. Similarly, the reason cannot be different effective masses because the masses are the same for electrons (or holes) in indirect and dark excitons. The second question deals with the polarization of the indirect exciton. Under circularly-polarized photoexcitation, the emission from the indirect exciton is  co-polarized with the photoexcitation. Given the nature of the indirect transition, as shown in Fig.~\ref{fig:scheme}(c), the question is why does the indirect exciton retain circular polarization? 

The organization of this paper is as follows.  We identify the zero-phonon resonance of the indirect exciton and explain its behavior in Sec.~\ref{sec:exp}. To explain the energy difference between the dark and indirect exciton, we study the  short-range electron-hole exchange interaction of excitons in Sec.~\ref{sec:eh_ex}. Using density functional theory (DFT) simulations for single-particle electronic states in conjunction with the stochastic variational method for the exciton envelope wavefunction, we show that the short-range electron-hole exchange interaction contributes a repulsive energy of $\sim$10~meV in the indirect exciton versus a vanishing contribution in the dark exciton case.  Section~\ref{sec:imp} includes analysis of the impurity-assisted recombination process of indirect excitons. Using symmetry analysis and numerical results, we show that the impurity-assisted recombination mechanism is dominated by spin-conserving intervalley transition of the electron to the hole's valley, whereas the opposite process is weaker (i.e., recombination mediated by transition of the hole to the electrons' valley). We use this finding to address the second question and explain why the emission of the indirect exciton retains the circular polarization of the photoexcitation.  Section~\ref{sec:imp} is a summary of the findings and the Appendices include further discussion on the emission spectra and technical details on DFT simulations and character tables. 
 
\section{The emission spectrum in (nearly) charge-neutral monolayer WS\lowercase{e}$_2$} \label{sec:exp}

\begin{figure}
\includegraphics[width=8.4cm]{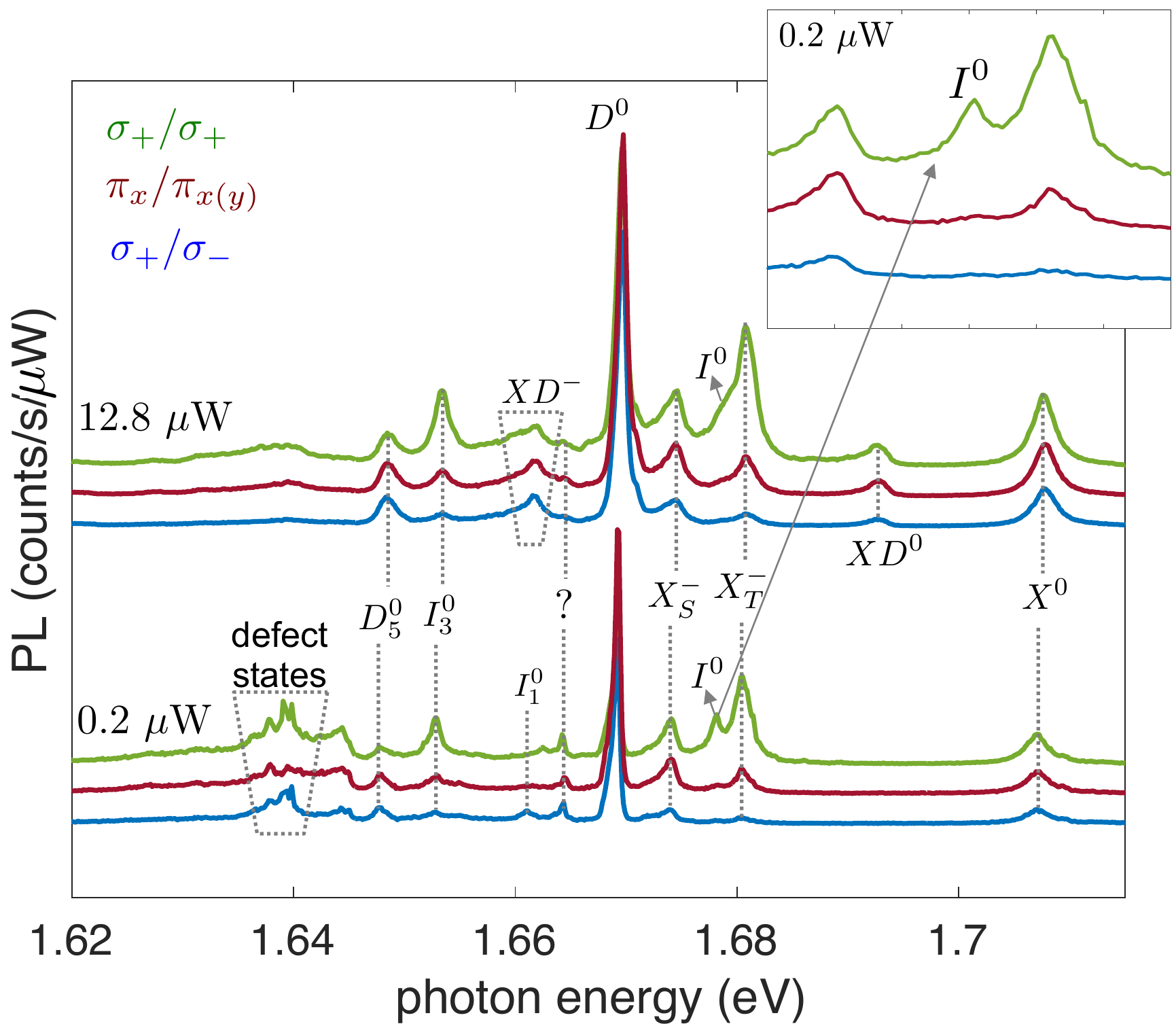}
\caption{ Emission spectrum of WSe$_2$ at 4~K when the monolayer is nearly charge neutral. The monolayer is photoexcited by HeNe laser (1.96~eV), focused on a spot size of $\sim$1~$\mu$m$^2$. The top curves show the spectra when the laser power is 12.8~$\mu$W and the bottom ones for 0.2~$\mu$W.  The emission intensity is normalized by the laser power in both cases and the spectra are shifted for clarity. The middle curve in each batch corresponds to co-polarized detection following linearly-polarized photoexcitation (the cross-polarized spectra are similar). The top/bottom curve in each batch corresponds to co-polarized/cross-polarized  detection following circularly-polarized photoexcitation.  Inset: The focus of this work is on the highlighted zero-phonon resonance of the  indirect exciton, $I^0$, which is seen best at low power excitation and co-polarized detection following circularly-polarized photoexcitation.  Appendix \ref{app:PL} includes further analysis and identification of all other labeled resonances.  \label{fig:exp}}
\end{figure}

Figure~\ref{fig:exp} shows the emission spectrum of WSe$_2$ at 4~K when the monolayer is nearly charge neutral. Details of the device and experiment are provided in Ref.~\cite{Robert_NatComm21}. The monolayer is photoexcited with HeNe laser at two laser powers, 0.2 and 12.8~$\mu$W, where the emission intensity is normalized by the laser power (see units of the $y$-axis). Three photoexcitation and detection configurations are shown for each of the two laser powers. The top and bottom curves in   each case show the spectrum following circularly polarized photoexcitation with co-polarized and cross-polarized detection, respectively. The middle curve in each case shows the results for linearly-polarized photoexcitation with co-polarized detection. Results for the cross-polarized detection under linearly-polarized photoexcitation are nearly identical to the co-polarized case (not shown) \cite{comment_pol}.  

The (zero-phonon) resonances of the bright, dark and indirect excitons are labeled in Fig.~\ref{fig:exp} by $X^0$, $D^0$, and $I^0$, respectively. The relatively strong emission intensity of  $D^0$ is a result of the large population of dark excitons (ground state). The dynamics between different exciton species determines their lifetimes ranging from a few ps in the case of bright excitons up to a ns when dealing with dark excitons in WSe$_2$ monolayer  \cite{Robert_PRB17,Singh_PRL16}.   The resonance $I^0$ emerges slightly below 1.68~eV  and is best seen at low power excitation. The salient feature of $I^0$ when the laser is circularly polarized compared with the linearly polarized case is consistent with the aforementioned enhanced lifetime of indirect excitons when their exchange-scattering relaxation to dark excitons is hampered by valley pumping \cite{Yang_PRB22}. Furthermore, comparing the circular polarization degrees of $I^0$ and $X^0$, extracted from the difference of co-circular and cross-circular polarized spectra divided by their sum, the emission from indirect excitons shows larger circular polarization degree than from bright excitons.

The emergence of the indirect exciton peak at weak photoexcitation is explained as follow.  Given the large mismatch between the crystal momentum of the electron and hole  in the indirect exciton, its zero-phonon radiative recombination can be mediated by scattering off the short-range potential of impurities. The impurities break the translation symmetry of the monolayer, thereby alleviating the need to conserve crystal momentum.  The interaction with impurities is optimal when the monolayer is nearly charge neutral and the laser power is relatively small. The impurities are then unscreened, the lifetime of the exciton is relatively long, and each exciton `sees' many impurities from which it can scatter. The presence of impurities is corroborated in Fig.~\ref{fig:exp} through the series of resonances around 1.64~eV when the laser power is 0.2~$\mu$W. These resonances correspond to strongly localized excitons \cite{Rivera_NatComm21}, where similar to $I^0$, their relative emission intensities diminish with respect to other resonances when the laser power is large. Appendix \ref{app:PL} includes a detailed discussion on the identity of all other labeled resonances in Fig.~\ref{fig:exp}. Below, we continue to analyze the indirect exciton. 

\section{Short-range electron-hole exchange interaction of excitons} \label{sec:eh_ex}

The electron-hole (\textit{e-h}) exchange interaction is effective when the electron in the conduction band and missing electron in the valence band have similar spins. As a result,  this interaction affects the binding energy of bright and indirect excitons while being ineffective in the case of dark excitons. In addition, the amplitude of the \textit{e-h} exchange interaction is commensurate with the spatial overlap between the electron and hole. The overlap is enhanced when a two-dimensional (2D) semiconductor is embedded in environments with reduced dielectric screening. In monolayer transition-metal dichalcogenides, for example, the long-range component of the \textit{e-h} exchange interaction is responsible for the fast valley depolarization of bright excitons  \cite{Yu_NatComm14,Yu_PRB14,Glazov_PSSB15,Yang_PRB20}.

The short-range  \textit{e-h} exchange interaction is known to be one of three factors that determine the energy difference between the resonances of bright and dark excitons in the emission spectrum (40-42~meV) \cite{Yang_PRB22}. The other two factors stem from the spin-orbit interaction. First, the electron effective mass in the bottom spin-split valley of the conduction band is  $\sim30\%$ heavier than that of an electron in the top valley. That is, the dark exciton has larger binding energy because it is heavier. Secondly,  the reference band-gap energy of the dark exciton is smaller by $\Delta_c$ due to the energy difference between the bottom and top spin-split valleys ($\Delta_c \sim 14$~meV in monolayer WSe$_2$ \cite{Kapuscinski_CommPhys21,Yang_PRB22}). 

The \textit{e-h} exchange energy of the indirect exciton is also believed to be responsible for the 10~meV splitting between $I_0$ and $D_0$ in WSe$_2$ \cite{He_NatComm20}. In this section, we briefly introduce the general theory of short-range \textit{e-h} exchange interaction developed by Bir and Pikus \cite{Pikus_Bir_71}, followed by numerical calculations of the short-range exchange energies of indirect, dark and bright excitons. 

\subsection{General theory}

The \textit{e-h} exchange interaction is a direct consequence of the Pauli exclusion principle, which results in antisymmetric form of the two-particle wavefunction
\begin{align}
\Psi = \frac{F(\bm{r}_1,\bm{r}_2)}{\sqrt{2}} 
[\psi_{\bm{k}_c}(\bm{r}_1)\psi_{\bm{k}_v}(\bm{r}_2) - \psi_{\bm{k}_c}(\bm{r}_2)\psi_{\bm{k}_v}(\bm{r}_1)].
\label{eq:Psi_antisym}
\end{align}
$\psi_{\bm{k}_c}(\bm{r})$  and $\psi_{\bm{k}_v}(\bm{r})$ are single particle  wavefunctions at the conduction and valence band extrema $\bm{k}_c$ and $\bm{k}_v$, respectively, with the form
\begin{align}
\psi_{\bm{k}}(\bm{r}) = e^{i\bm{k}\cdot\bm{r}}\phi_{\bm{k}}(\bm{r}),
\label{eq:psi}
\end{align}
where $\phi_{\bm{k}}(\bm{r})$ is the normalized Bloch function with period of the crystal lattice.
The exciton envelope function is symmetric and depends on the relative position of the electron and hole, $F(\bm{r}_1,\bm{r}_2)=F(\bm{r}=\bm{r}_1-\bm{r}_2)$. In many practical cases, the exciton ground state can be approximated by the 1$s$ wavefunction of an effective hydrogen model.

The \textit{e-h} exchange energy of the exciton is written as
\begin{align}
E_\text{X} = |F(r=0)|^2 I,  
\label{eq:Ex}
\end{align}
originating from the cross integral contribution [between the two terms in Eq.~(\ref{eq:Psi_antisym})] in the matrix element of the Coulomb interaction 
\begin{align}
I & = -\langle\psi_{\bm{k}_c}(\bm{r}_1)
\psi_{\bm{k}_v}(\bm{r}_2) 
|U(r)| 
\psi_{\bm{k}_c}(\bm{r}_2)
\psi_{\bm{k}_v}(\bm{r}_1)\rangle
\label{eq:exchange_integral}\\
& = -\langle\phi_{\bm{k}_c}(\bm{r}_1)
\phi_{\bm{k}_v}(\bm{r}_2) 
|e^{i(\bm{k}_v-\bm{k}_c)\bm{r}}U(r)| 
\phi_{\bm{k}_c}(\bm{r}_2)
\phi_{\bm{k}_v}(\bm{r}_1)\rangle\nonumber,
\end{align}
where $U(r)$ is the Coulomb potential.

To examine the integration in Eq.~(\ref{eq:exchange_integral}), we follow the procedure by Pikus and Bir \cite{Pikus_Bir_71}, and consider the Fourier series of the product between conduction and valence Bloch functions,
\begin{align}
\phi_{\bm{k}_c}^*(\bm{r})\phi_{\bm{k}_v}(\bm{r}) = \frac{1}{\Omega}\sum_{j} B_j(\bm{k}_c; \bm{k}_v) e^{i\bm{b}_j\cdot\bm{r}}.
\label{eq:Bj_kc_kv}
\end{align}
$\Omega$ is the monolayer volume and $\bm{b}_j$ is the $j$-th reciprocal lattice vector. The  Fourier coefficients $B_j(\bm{k}_c; \bm{k}_v)$ have two useful properties. First, for a direct-gap exciton with $\bm{k}_c = \bm{k}_v$, $B_j$ vanishes for the long-range component with $\bm{b}_j = 0$ due to orthogonality of the conduction and valence states. Secondly, by comparing Eq.~(\ref{eq:Bj_kc_kv}) with its complex conjugation, one gets the relation
\begin{align}
B_j^*(\bm{k}_c; \bm{k}_v) = B_{-j}(\bm{k}_v; \bm{k}_c).
\label{eq:Bj_kc_kv1}
\end{align}
Substituting Eqs.~(\ref{eq:Bj_kc_kv}) and (\ref{eq:Bj_kc_kv1}) in Eqs.~(\ref{eq:Ex}) and (\ref{eq:exchange_integral}), and then integrating over $\bm{r}_1$ and $\bm{r}_2$, we get
\begin{align}
E_\text{X} = -|F(0)|^2\sum_{j} |B_j(\bm{k}_c; \bm{k}_v)|^2 U_{\bm{k}_c - \bm{k}_v - \bm{b}_j}.
\label{eq:exchange_energy}
\end{align}
$U_{\bm{k}_c - \bm{k}_v - \bm{b}_j}$ is the inverse Fourier transform of the Coulomb potential, given by 
\begin{align}
U_{\bm{q}} = \frac{1}{\Omega}\int U(\bm{r}) e^{-i\bm{q}\cdot\bm{r}}d\bm{r}.
\label{eq:inv_fourier_V}
\end{align}
 
Eq.~(\ref{eq:exchange_energy}) points to the `short-range' nature of this exchange energy.  For direct-gap excitons, $\bm{k}_c=\bm{k}_v$, the long-range term $U_{\bm{q}\sim 0}$ is excluded by the diminished $B_j$ when $\bm{b}_j = 0$. For indirect excitons with large difference between $\bm{k}_c$ and $\bm{k}_v$, the term $U_{\bm{q}\sim 0}$ is not part of the sum. The outcome is that the short-range exchange energy is expected to be similar in direct and indirect excitons. 

\subsection{Numerical calculation}
The exchange energies for different types of excitons is calculated using Eq.~(\ref{eq:exchange_energy}). We discuss the numerical procedure for evaluating the three ingredients of this equation: the Coulomb potential $U_{\bm{q}}$, the Fourier coefficients $B_j$, and the square amplitude of the exciton envelope wavefunction at the origin, $|F(0)|^2$.  

The Coulomb interaction in our quasi-2D system is subjected to dielectric screening in the monolayer and its surrounding environment. The Coulomb potential in the long-wavelength limit ($q\rightarrow 0$) resembles the Rytova-Keldysh potential in which the dielectric constant is linear in $q$ \cite{Cudazzo_PRB11,Rytova_PMPA67,Keldysh_JETP79}. In the shortwave limit, on the other hand, ab-initio calculations show that the Coulomb potential acquires the three-dimensional potential form when the wavenumber is comparable or larger than the reciprocal vector close to the edge of the first Brillouin zone \cite{Latini_PRB15,Qiu_PRB16}. With this consideration, we capture the asymptotic trend of the quasi-2D Coulomb potential toward the 3D limit at large wavevectors
\begin{align}
U_{\bm{q}} = -\frac{1}{\Omega}
\frac{e^2}{\epsilon'\epsilon_0} \frac{1}{q^2},
\label{eq:coulomb}
\end{align}
where $\epsilon'$ is an `effective' relative dielectric constant. Intuitively, the dielectric screening effect fades at extremes shortwaves, $\epsilon' \rightarrow 1$ when $q \rightarrow \infty$, due to lack of background charge between two charges whose distance from each other is much smaller than the lattice constant.

The unit-less Fourier coefficients $B_j$ are obtained by convolution of conduction and valence Bloch functions, 
\begin{align}
\phi_{\bm{k}_{c,v}}(\bm{r}) &= \frac{1}{\sqrt{\Omega}} \sum_{\bm{b}_j} C_j^{\bm{k}_{c,v}} e^{i\bm{b}_j\cdot\bm{r}},
\label{eq:phi}
\end{align}
which yields
\begin{align}
B_j(\bm{k}_c; \bm{k}_v) = \sum_{m, n} {C_m^{\bm{k}_c}}^*C_n^{\bm{k}_v}\delta_{\bm{b}_n-\bm{b}_m, \bm{b}_j}.
\label{eq:Bj_kc_kv_1}
\end{align}
The plane-wave representation of the eigenstates in Eq.~(\ref{eq:phi}) is calculated using the \textit{ab initio} package {\sc QUANTUM ESPRESSO} \cite{QE}. Details of the calculation are provided in Appendix~\ref{appendix:DFT}.

The unit-less exciton envelope wavefunction, according to its definition in Eq.~(\ref{eq:Psi_antisym}), can be approximated by decomposition into the in-plane and out-of-plane components, 
\begin{align}
|F(0)|^2 = \Omega \left| \Phi_{1s}^{2D}(0) \right|^2 \left |\Phi_{g}^{z}(0)\right|^2 = \left(\frac{2}{\pi}\right)^{\frac{3}{2}}\frac{\Omega}{a_B^2 a_z}.
\label{eq:F0}
\end{align}
$\Phi_{1s}^{2D}(\bm{r}_\parallel)$ is the $1s$ state of the exciton in the monolayer plane, where $a_B = \sqrt{\frac{2}{\pi}}[\Phi_{1s}^{2D}(r_\parallel=0)]^{-1}$ is its effective Bohr radius. Different from the short-range \textit{e-h} exchange interaction, it is the long-range component of the effective two-dimensional Coulomb interaction that governs the exciton binding energy and wavefunction. The explicit form of the Coulomb interaction is the Rytova-Keldysh potential in wavenumber space \cite{Rytova_PMPA67,Keldysh_JETP79,Cudazzo_PRB11},
\begin{align}
U_{RK}(\bm{q}) = -\frac{1}{A}\frac{e^2}{2\epsilon\epsilon_0} \frac{1}{q_\parallel(1+r_0q_\parallel)}.
\label{eq:RK}
\end{align}
$q_\parallel$ is the in-plane component of the wavevector, $r_0 = 1.18$~nm is the effective polarizability distance of the monolayer, and $\epsilon = 3.8$ is the dielectric constant of the encapsulating layers (hexagonal boron nitride) \cite{VanTuan_PRB18}.  Using the Rytova-Keldysh potential and effective masses of the electron and hole at their respective band edges (Table~\ref{tab:I}), $\Phi_{1s}^{2D}(\bm{r}_\parallel)$ is calculated by the stochastic variational method \cite{Varga_PRC95,VanTuan_PRB18}. The extracted Bohr radii in WSe$_2$ and WS$_2$ monolayers are given in Table~\ref{tab:I}. 

\begin{table} 
\caption{Calculated values of the \textit{e-h} short-range exchange energy ($E_X$) of three types of excitons ($D_0$, $I_0$, and $X_0$), together with the parameters used in the calculations. The first (second) value in each entry corresponds WSe$_2$ (WS$_2$).}
\label{tab:I}
\renewcommand{\arraystretch}{1.5}
\begin{tabular}{p{20mm}|c|c|c}
\hline 
WSe$_2$, WS$_2$& $D_0$ & $I_0$ & $X_0$ \\ \hline \hline
$m_e$ [$m_0$] & \multicolumn{2}{c|}{0.4, 0.36} &  0.29, 0.27  \\\hline
$m_h$ [$m_0$] & \multicolumn{3}{c}{0.36, 0.36} \\\hline
$a_B$ [\AA] & \multicolumn{2}{c|}{17.7, 18.3} & 19.6, 20.1\\\hline
$a_z$ [\AA]  & \multicolumn{3}{c}{4.7, 5.2} \\\hline
$\epsilon'$ & \multicolumn{3}{c}{2.9, 2.9} \\\hline
$S_X$ [meV$\cdot$ \AA$^3$]  &  3.5, 3.0  &  33.0, 27.7  &  40.9, 34.9   \\
\hline
$E_X$ [meV]  & $\quad$ 1.2, 0.9 $\quad$ & $\quad$ 11.4, 8.4 $\quad$ & $\quad$ 11.5, 8.1 $\quad$  \\
\hline
\end{tabular}
\end{table}

Lastly, $|\Phi_{g}^{z}(0)|^2 = \sqrt{\frac{2}{\pi}}\frac{1}{a_z}$ is the square amplitude of the out-of-plane envelope function at the origin. The values of $a_z$ (Table~\ref{tab:I}) are the sum of out-of-plane extensions of the electron and hole Bloch wavefunctions calculated from {\sc QUANTUM ESPRESSO} package (see Appendix~\ref{appendix:DFT}).
This approximation is based on the consideration that edge states of the conduction and valence bands are dominated by $d$-orbitals of tungsten atoms in the mid-plane of the monolayer \cite{Liu_PRB13}. Since the thickness of the monolayer (distance between chalcogen sublayers) is only a fraction of $a_B$, the out-of-plane motion of the electron and hole is dominated by the monolayer confinement, whereas the influence of their Coulomb attraction is secondary. To lowest order, the out-of-plane confinement is approximated by a 1D quadratic quantum well with Gaussian ground state characterized by $a_z$, whose value is comparable to twice the thickness of the monolayer.

With all these considerations, the exciton exchange energy Eq.~(\ref{eq:exchange_energy}) is reduced to a simpler form
\begin{align}
E_\text{X} =  \left(\frac{2}{\pi}\right)^\frac{3}{2} \frac{S_\text{X}}{a_B^2 a_z},
\label{eq:exchange_energy_1}
\end{align}
with
\begin{align}
S_\text{X} = \frac{e^2}{\epsilon'\epsilon_0} \sum_{j} \frac{|B_j(\bm{k}_c; \bm{k}_v)|^2}{|\bm{k}_c - \bm{k}_v - \bm{b}_j|^2}.
\label{eq:sum_BU}
\end{align}
The calculated $S_\text{X}$ and $E_\text{X}$  of indirect, dark, and bright excitons are given in Table~\ref{tab:I} for both WSe$_2$ and WS$_2$. To match the energy splitting between the resonances $I_0$ and $D_0$ in WSe$_2$  ($\sim10$~meV),  we choose $\epsilon' = 2.9$. This value is consistent with ab-initio calculations of the dielectric function in the limit that the wavenumber is comparable to the reciprocal vector close to the edge of the first Brillouin zone \cite{Latini_PRB15,Qiu_PRB16}. Regarding monolayer WS$_2$, there is no available experimental result yet (to the best of our knowledge), which is fine enough to resolve the resonance $I_0$. Our calculation shows that the difference between the two materials is marginal, consistent with the fact that the electron and hole states in both materials are governed by atomic orbitals of tungsten atoms. 

The calculated short-range \textit{e-h} exchange energies of indirect and the bright excitons are similar for both WS$_2$ and WSe$_2$, $E_X \simeq 10$~meV. Indeed, the scale of $E_X$ should be in the ballpark of $\sim E_0^2/E_g$  \cite{Pikus_Bir_71}, where $E_0 \sim 200$~meV is the exciton binding energy and $E_g \sim 2$~eV is the energy band gap of free \textit{e-h} pair transitions. For dark excitons, the opposite majority spin components of their electron and hole states suggest that the Fourier coefficients $B_j$ in Eq.~(\ref{eq:exchange_energy}) should vanish to lowest order \cite{Yu_PRB14}. Indeed, the calculated short-range \textit{e-h} exchange energy of the dark exciton is about one order of magnitude smaller than those of bright and indirect excitons ($E_X \sim 1$~meV).  It is not completely vanished because of spin-orbit coupling to remote (high-energy) conduction bands with opposite spin \cite{Song_PRL13,Slobodeniuk_2DMater16,Wang_PRL17}. This coupling mediates the recombination of dark excitons via the out-of-plane dipole moment \cite{Tang_NatComm19,Liu_PRL19}. Using the calculated Bloch wavefunctions, the small spin-minority mixing of the electron in the dark exciton state is around 4\% for both WSe$_2$ and WS$_2$.  

\section{Impurity-assisted recombination} \label{sec:imp}
This section deals with the zero-phonon recombination process of indirect excitons, which underlies the resonance $I^0$. As discussed in Sec.~\ref{sec:exp}, the presence of impurities is needed to alleviate the large mismatch between the wavevectors (crystal momenta) of the electron and hole during recombination of the indirect exciton. The large wavevector mismatch suggests that the scattering is from the short-range part of the impurity potential.  The role of (elastic) scattering off impurities during recombination  is equivalent to the role of the exciton-phonon interaction when dealing with phonon-assisted recombination processes. In the latter case, the emission (or absorption) of shortwave phonons is required to accommodate the large wavevector mismatch \cite{Li_PRL10}.

It is emphasized that the resonance $I^0$ stems from scattering of delocalized indirect excitons off impurities rather than recombination of localized indirect excitons that are bound to the impurity. If the latter was the case, then we would expect to see the resonance at lower energy.  Yet, there is a fundamental relation between the scattered (delocalized) and bound (localized) states of an impurity potential \cite{Gantmakher_Levinson_Book,Song_PRL14}. That is, an impurity potential that induces large binding energy (localization)  brings in stronger short-range scattering.

\subsection{Second order perturbation and selection rules}

\begin{figure}
\includegraphics[width=8.5cm]{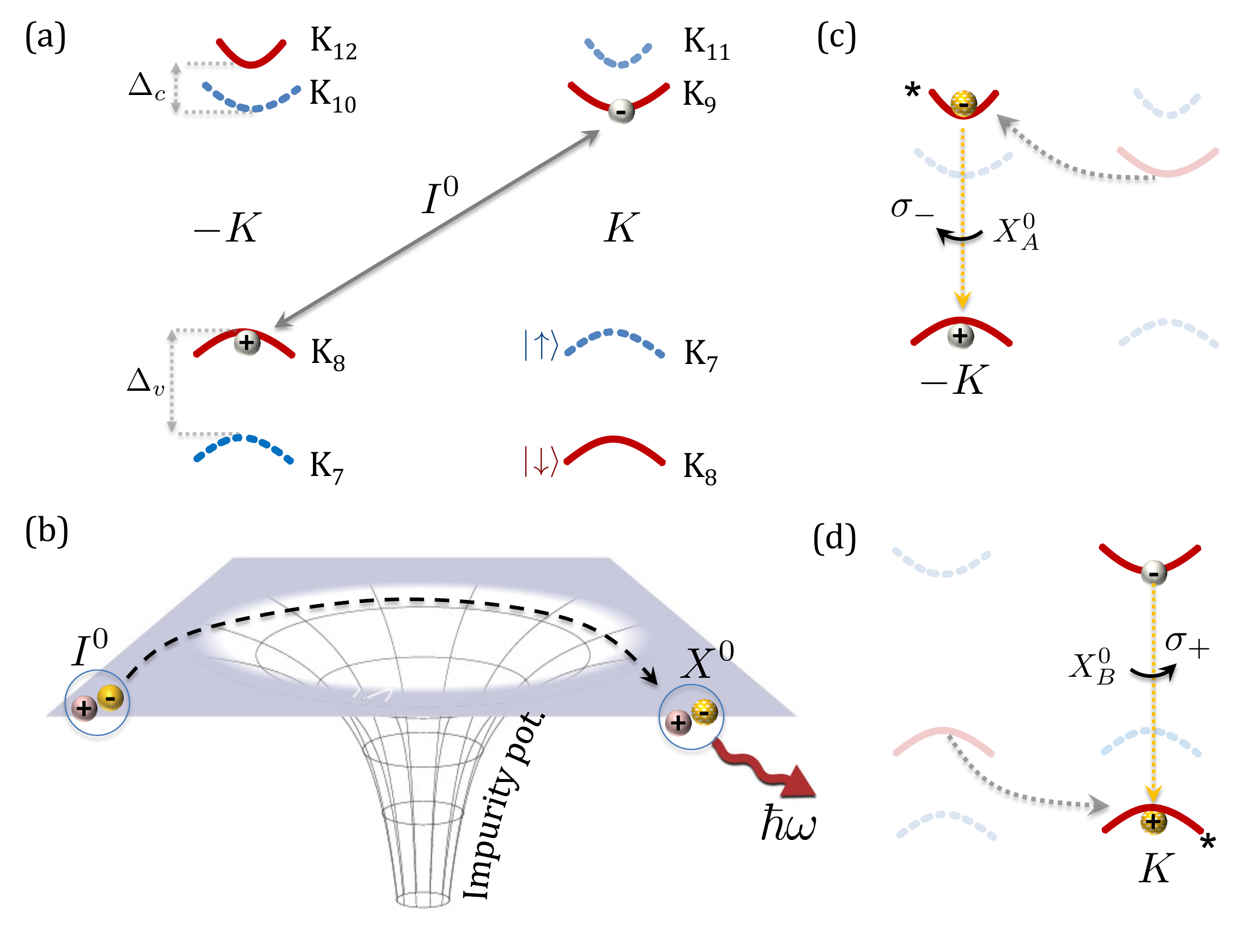}
\caption{Impurity-assisted recombination of indirect excitons. (a) $\mathbf{k}$-space representation of the indirect exciton. $\Delta_{c(v)}$ is the spin-splitting energy in the conduction (valence) band. $K_{7-12}$ are irreducible representations of the  $C_{3h}$ double point group (Appendix ~\ref{appendix:B}), representing transformation properties of electronic states in the corresponding energy bands \cite{Song_PRL13}.  (b) Spin-conserving intervalley scattering of the indirect exciton off the short-range potential of the impurity, after which the exciton becomes virtually bright and emits light. (c)/(d) Photon emission from the virtual bright exciton state following intervalley transition of the electron/hole.  The helicity of the emitted photon  is opposite in the two cases. }
\label{fig:recom}
\end{figure}

A straightforward picture of the recombination process can be described as a second-order transition, in which scattering of the indirect exciton off the impurity induces intervalley transition to an intermediate bright exciton state. The ensuing radiative recombination from the (virtual) bright exciton emits a photon according to the optical selection rules of bright excitons.  Figure~\ref{fig:recom} shows the steps of the recombination process. The indirect exciton in the initial state, shown in Fig.~\ref{fig:recom}(a), experiences spin-conserving intervalley transition to intermediate bright exciton state following scattering off the short-range part of the impurity potential, as shown in Fig.~\ref{fig:recom}(b). The transition to bright exciton can be governed by scattering of the electron or hole component, as shown in Fig.~\ref{fig:recom}(c) and (d), respectively.  When the hole goes through the intervalley transition, the emitted photon comes from dipole transition of type-B excitons (i.e., it involves the bottom spin-split valley of the valence band).  The bright exciton state is a virtual intermediate step during recombination of indirect excitons due to the energy mismatch between the indirect and bright excitons. 

Using second-order perturbation theory, the zero-phonon recombination rate of indirect exciton with translation wavevector $\mathbf{k}$ is \cite{Cardona_book,VanTuan_PRL19}
\begin{eqnarray}
\frac{1}{\tau_{I^0_{\mathbf{k}}}} &=& N_i \, \frac{2\pi }{\hbar} \,\sum_{\mathbf{Q},j} \delta\left( \hbar \omega_{I^0_{\mathbf{k}}} - \hbar c Q \right) \nonumber \\
&\cdot& \left|\sum_{n,\mathbf{q}}\frac{ \langle  0 \,|\, \hat{H}_\text{LM}(\mathbf{Q},j)\,|\, X^0_{n,{\mathbf{q}}} \rangle        \langle X^0_{n,{\mathbf{q}}} \,|\,\hat{U}_{\text{imp}}\,|\, I^0_{\mathbf{k}} \rangle    }{ \hbar \omega_{I^0_{\mathbf{k}}} - \hbar \omega_{X^0_{n,{\mathbf{q}}}}      }\right|^2\!\!\!. \,\,\,\,\,\,\,\,\,\,\,
\label{eq:2nd_order}
\end{eqnarray}
$N_i = An_i$ is the number of impurities in the monolayer, where $A$ is the monolayer area and $n_i$ is the impurity density. The final state of the recombination process is the ground-state of the monolayer (filled valence band with no excitons) and emitted photon whose three-dimensional wavevector is $\mathbf{Q}=\{\mathbf{Q}_{\parallel},Q_z\}$, its polarization state is denoted by $j$, and its energy is $ \hbar c Q\sim$1.68~eV according to the emission spectrum. The initial state of the recombination process is the indirect exciton whose energy is $\hbar \omega_{I^0_{\mathbf{k}}}=E_g - \varepsilon_{I^0} + \hbar^2 k^2/2(m_e+m_h)$, where $E_g$ is the band gap energy, $\varepsilon_{I^0}$ is the binding energy, and the last term is its (small) kinetic energy. 

The transition probability in the second line of Eq.~(\ref{eq:2nd_order}) is governed by interference from transition paths coming from the sum over intermediate direct-gap exciton states. The sum includes the state of the direct-gap exciton ($n$), and its in-plane translational wavevector ($\mathbf{q}$). The first matrix element corresponds to light-matter interaction of  the intermediate direct-gap exciton, and the second matrix element to impurity-assisted intervalley scattering needed to turn the exciton from indirect (initial state) to direct (intermediate state). In principle, the interference effect is accounted for by letting $n$ run over all sorts of intermediate direct-gap excitons states, including bright and dark excitons, type-B excitons, and direct-gap exciton states that involve remote energy bands. In practice, however, we should only account for the nearby bright exciton. The reason for neglecting dark excitons is that that their dipole transition is much smaller, $\langle  0 \,|\, \hat{H}_\text{LM}(\mathbf{Q},j)\,|\, D^0_{n,{\mathbf{q}}} \rangle \rightarrow 0$.  The reasons for neglecting other bright exciton states is their large energy  detuning energy from the indirect exciton. For example,  $\hbar \omega_{I^0_{\mathbf{k}}} - \hbar \omega_{X^0_{n,{\mathbf{q}}}} \sim$30~meV when $n$ refers to the ground state of the bright exciton compared with $\sim$160~meV when considering its 2$s$ state (whose dipole moment is also weaker than that of 1$s$). In addition, the energy detuning is much smaller than that of type-B excitons ($\gtrsim$400~meV), governed by the relatively large spin-splitting energy in the valence band.

Further support for neglecting intermediate type-B bright excitons is by symmetry considerations. According to the method of invariances \cite{BirPiKus_Book}, whether a matrix element is allowed or forbidden depends on the symmetries of the bra and ket states and the interaction. Specifically, given the irreducible representations (IRs) of the bra and ket, a nonzero matrix element requires that their direct product includes the IR of the interaction. 
In our case, the symmetry of the $K$ and $K'$ points of the Brillouin zone is governed by a space group whose little group is the point group $C_{3h}$. 
With spin involved, the double group of $C_{3h}$ includes twelve IRs (Appendix~\ref{appendix:B}), among which $K_7\sim K_{12}$ are associated with physical states, as labeled in Fig.~\ref{fig:recom}(a)  \cite{Song_PRL13}. The selection rule for spin-conserving intervalley scattering of the electron component in the indirect exciton yields
\begin{align}
&K_9^* \times K_{12} = K_3\,, \label{eq:selection_rules_K3}
\end{align}
and of the hole component yields
\begin{align}
K_8^* \times K_8 = K_1\,.  \label{eq:selection_rules_K1}
\end{align}
Unless the impurity is magnetic, spin-flip intervalley transition between indirect and dark excitons is forbidden by time-reversal symmetry (for either electron or hole transitions).  The symmetry of the intervalley scattering potential $\hat{U}_{\text{imp}}$ is dictated by its short-range nature, such that contributions come only from Fourier components with large wavevector components to compensate the crystal momentum mismatch of the electron and hole $2K = K - (-K)$ [see Eq.~(\ref{eq:inv_fourier_V})]. Therefore, the short-range part of $\hat{U}_{\text{imp}}$ transforms as in-plane vector which is compatible with the IR $K_3$ \cite{Song_PRL13}. According to Eq.~(\ref{eq:selection_rules_K3}), the spin-conserving intervalley transition should involve the electron component, governed by the mechanism depicted in Fig.~\ref{fig:recom}(c). 

The calculation of the recombination rate  in Eq.~(\ref{eq:2nd_order}) can be further simplified using the following viable approximations. The first approximation is to replace the spin-conserving intervalley matrix element of the exciton with that of a free electron,
\begin{align}
 M_\text{imp} = \langle X^0_{{\mathbf{q}}} \,|\,\hat{U}_{\text{imp}}\,|\, I^0_{\mathbf{k}} \rangle \approx  \langle\psi_{K_c,s}|\hat{U}_\text{imp}|\psi_{-K_c,s}\rangle , \label{eq:Mimp}
\end{align}
where $s$ is the spin index, and $| \psi_{\pm K_c,s} \rangle$ are the conduction-band edge states at the $\pm K$ valleys. The approximation relies on the argument that the hole is a spectator during the short-range scattering  of the electron, which is valid when $2K*a_B \gg 1$. This condition signifies that the short-range impurity scattering is due to central-cell correction of the lattice potential within a unit cell, whereas the exciton Bohr radius extends over several unit cells. In addition, we can safely neglect the small wavevector components of the indirect and bright exciton ($k,q \ll K$), and consider the transition between conduction states exactly at  $K$ and $-K$. The reason is that the selection rule in Eq.~(\ref{eq:selection_rules_K3}) is of zeroth-order, meaning that $M_\text{imp}$ is largely independent of the relatively small translational wavevectors of the indirect and bright excitons.

The last approximation we make is to assume that the intermediate bright exciton is delocalized, and therefore conserve the crystal momentum during the optical transition.  The light matter interaction matrix element becomes \cite{Wang_PRB14}, 
\begin{eqnarray}
 \langle  0 \,|\, \hat{H}_\text{LM}(\mathbf{Q},j)\,|\, X^0_{\mathbf{q}} \rangle    =   \frac{e v_0}{a_B} \sqrt{\frac{\hbar }{\pi \epsilon_0 \epsilon_b c } \cdot \frac{1}{Q} \cdot  \frac{A}{V}}\,\,,
 \label{eq:lm}
\end{eqnarray}
where $\epsilon_b$ is the dielectric constant of the encapsulating barriers, and $v_0 \sim 5 \cdot 10^7$~cm/s comes from the optically-active dipole transition between conduction and valence band states \cite{Scharf_JPCM19}. $V$ is the volume of the monolayer and the surrounding environment, resulting from quantization of the electromagnetic field ($\Omega = Aa_z$ where $a_z$ is the effective monolayer thickness whereas $V=AL_z$).   Putting everything together,  the zero-phonon recombination rate of indirect excitons can be approximeated as
\begin{eqnarray}
\frac{1}{\tau_{I^0}} \approx \frac{ | M_\text{imp} |^2 }{\Delta_{ix}^2} \, \frac{A^2 n_i}{\pi^2 a_B^2} \, \left( \frac{v_0}{c}\right)^2 \, \frac{e^2 q_0}{\hbar \epsilon_0 \epsilon_b}.
 \label{eq:rate}
\end{eqnarray}
$q_0 = \omega_{I^0}/c \approx 0.85~\mu$m$^{-1}$ and $\Delta_{ix} \approx 30$~meV is the energy difference between the resonances of $I^0$ and $X^0$.

\subsection{Numerical calculation of $M_\text{imp}$}

Given the large momentum mismatch involved in the intervalley transition, the impurity potential $\hat{U}_\text{imp}$ is modeled through the three-dimensional limit of the bare Coulomb potential in Eq.~(\ref{eq:coulomb}), where as before $\epsilon' = 2.9$. The intervalley matrix element in Eq.~(\ref{eq:Mimp}) can then be written as
\begin{align}
M_\text{imp} = \sum_j \! B_j(\bm{K}_c,s; -\bm{K}_c,s) U_{2\bm{K}-\bm{b}_j} \equiv \frac{\mathcal{C}_{\text{imp}}}{\Omega}, \label{eq:2Mimp}
\end{align}
where the unit-less Fourier coefficients $B_j$ are calculated from Eqs.~(\ref{eq:phi})-(\ref{eq:Bj_kc_kv_1}) using same-spin conduction Bloch functions at $\pm K$ points. Similar to the analysis of short-range exchange interaction, the planewave representation of these two states is provided by {\sc QUANTUM ESPRESSO} \cite{QE}. The calculated values of $\mathcal{C}_{\text{imp}}$ are 0.48~eV$\cdot$nm$^3$ and 0.36~eV$\cdot$nm$^3$ in WSe$_2$ and WS$_2$ monolayers, respectively. Substituting these values in Eq.~(\ref{eq:rate}) and assuming impurity density of $n_i$$\,=\,$10$^{11}$~cm$^{-2}$, we get that $\tau_{I^0}$ is about 1.5~ns and 2~ns in WSe$_2$ and WS$_2$ monolayers, respectively. These recombination time scales are comparable to the lifetime of the dark exciton through pump-probe experiments of charge neutral monolayers. 

\section{Summary}  \label{sec:conc}
The analysis provided in this study sheds light on two important attributes of the indirect exciton. Through calculation of the short-range electron-hole exchange interaction, we have shown that the binding energy of indirect and bright excitons is weakened by about 10~meV in WSe$_2$ and WS$_2$ monolayers. This finding explains why the indirect exciton emerges about 10~meV above the dark exciton,  which is largely insensitive to the electron-hole exchange interaction. The second finding of this work is the calculation of impurity-assisted recombination of indirect excitons. We have shown that the recombination is dominated by intervalley scattering of the electron component in the indirect exciton, leading to light emission that follows the selection rule from the valley of the hole. Given that holes are more resilient in keeping their valley post photoexcitation with circularly polarized light (i.e., the indirect exciton is formed after the electron changes its valley \cite{He_NatComm20,Yang_PRB22}), the emission of the indirect exciton is co-circularly polarized if the electron finds a way back to the photoexcited valley.  The impurity-assisted recombination process supports this scenario, thereby explaining why the resonance $I^0$ in the emission spectrum retains the circular polarization of the laser. That the electron goes through intervalley scattering and the hole remains in the same valley is further supported by the sign of the $g$-factor of $I^0$ \cite{He_NatComm20}. The higher circular polarization degree from emission of the indirect exciton compared with that of the bright exciton can be explained by faster valley depolarization of bright excitons due to the long-range electron-hole exchange interaction \cite{Yu_NatComm14,Yu_PRB14,Glazov_PSSB15,Yang_PRB20}. The formalism we have developed  for impurity-assisted recombination of indirect excitons can be straightforwardly generalized to similar recombination processes from trions and other dark excitonic species. 

\acknowledgments{Work at the University of Rochester was supported by the Office of Naval Research, under Contract No. N000142112448 (Pengke Li) and the Department of Energy, Basic Energy Sciences, under Contract No. DE-SC0014349 (Dinh Van Tuan). Cedric Robert and Xavier Marie acknowledge funding from ANR ATOEMS and ANR MagicValley. Xavier Marie also acknowledges the Institut Universitaire de France.}
\appendix
\section{Review of the resonances in Fig.~\ref{fig:exp}} \label{app:PL}
\begin{figure}[t]
\includegraphics[width=8.4cm]{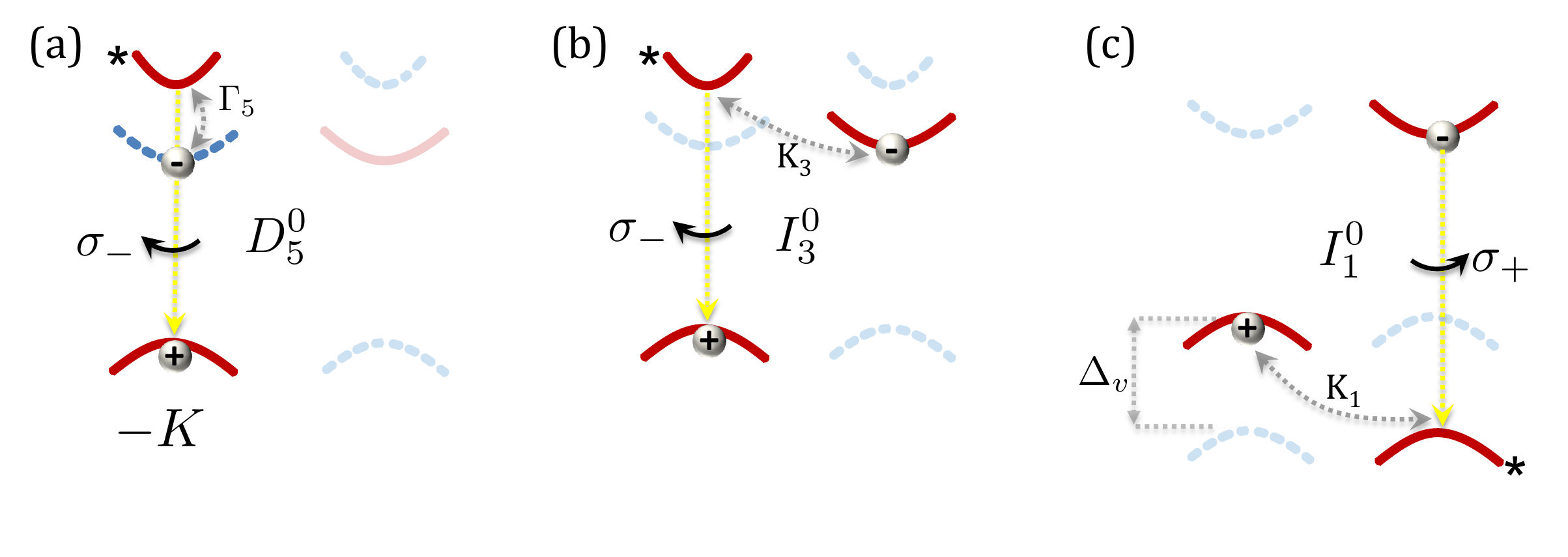}
\caption{Phonon-assisted recombination of the dark and indirect excitons. The highlighted energy bands correspond to electronic states that are involved in the recombination process, where the bands with an asterisk symbol correspond to the bright-exciton intermediate (virtual) state after phonon emission.  (a) Recombination of the dark exciton  assisted by  the zone-center $\Gamma_5$ phonon. (b)/(c) Recombination of the indirect exciton assisted by the zone-edge $K_{3/1}$ phonon.
\label{fig:phonon_replica}}
\end{figure}

The focus in the main paper was on the bare (zero-phonon) resonance of the indirect exciton ($I^0$). For completeness, we go over the rest of the resonances in Fig.~\ref{fig:exp}.
The phonon-assisted resonances of the dark exciton and indirect excitons are labeled by  $D^0_5$, $I^0_3$ and $I^0_1$ \cite{He_NatComm20,Yang_PRB22}. Their physical processes are depicted in Fig.~\ref{fig:phonon_replica}, and they involve the zone-center phonon $\Gamma_5$ and zone-edge phonons $K_1$ and $K_3$, as supported by symmetry arguments \cite{Song_PRL13,Dery_PRB15}. The phonon replica of the dark exciton, $D^0_5$, involves emission of the phonon mode $\Gamma_5$ whose energy is $\sim$21~meV.  The phonon-exciton interaction triggers an intravalley spin-flip transition of the electron, such that the optical transition involves the electronic state from the optically active valley of the conduction band (i.e., the bright exciton is an intermediate virtual state). As such, the valley of the dark exciton can be identified from the circular polarization of the resonance $D^0_5$.

The phonon replicas of the indirect exciton involve spin-conserving intervalley transition of the electron ($I^0_3$) or hole ($I^0_1$). The electron transition is assisted by emission of the zone-edge phonon $K_3$ whose energy is $\sim$26~meV, and as before it uses the bright exciton as an intermediate virtual state. The hole transition is assisted by emission of the zone-edge phonon $K_1$ whose energy is $\sim$17-18~meV, where the intermediate virtual state in this case is a type-B bright exciton which involves the bottom valley of the valence band. Figure~\ref{fig:exp} shows that the signature of the replica $K_3$  is much stronger than that of $K_1$ and that their polarizations is opposite. The reason for the dominance of $I^0_3$ is that the intermediate bright exciton state in this case is much closer to the energy of the indirect exciton. The opposite polarization is reasoned by the fact that the intermediate bright exciton is from opposite valleys. Finally, we notice that $I^0_3$ is co-polarized whereas $I^0_1$ is cross-polarized with respect to the circular polarization of the laser. This behavior indicates a significant relaxation of photoexcited bright excitons to indirect ones through emission of $K_3$ phonons, during which the hole is kept in the photoexcited valley. Note that the energy relaxation from bright to indirect exciton is not part of the recombination process of $I^0_3$. The phonon-assisted recombination of the indirect exciton through the mode $K_3$ comes at a later phase when the indirect excitons (i.e., the recombination involves emission of a different $K_3$ phonon).   

\begin{figure}[t]
\includegraphics[width=7cm]{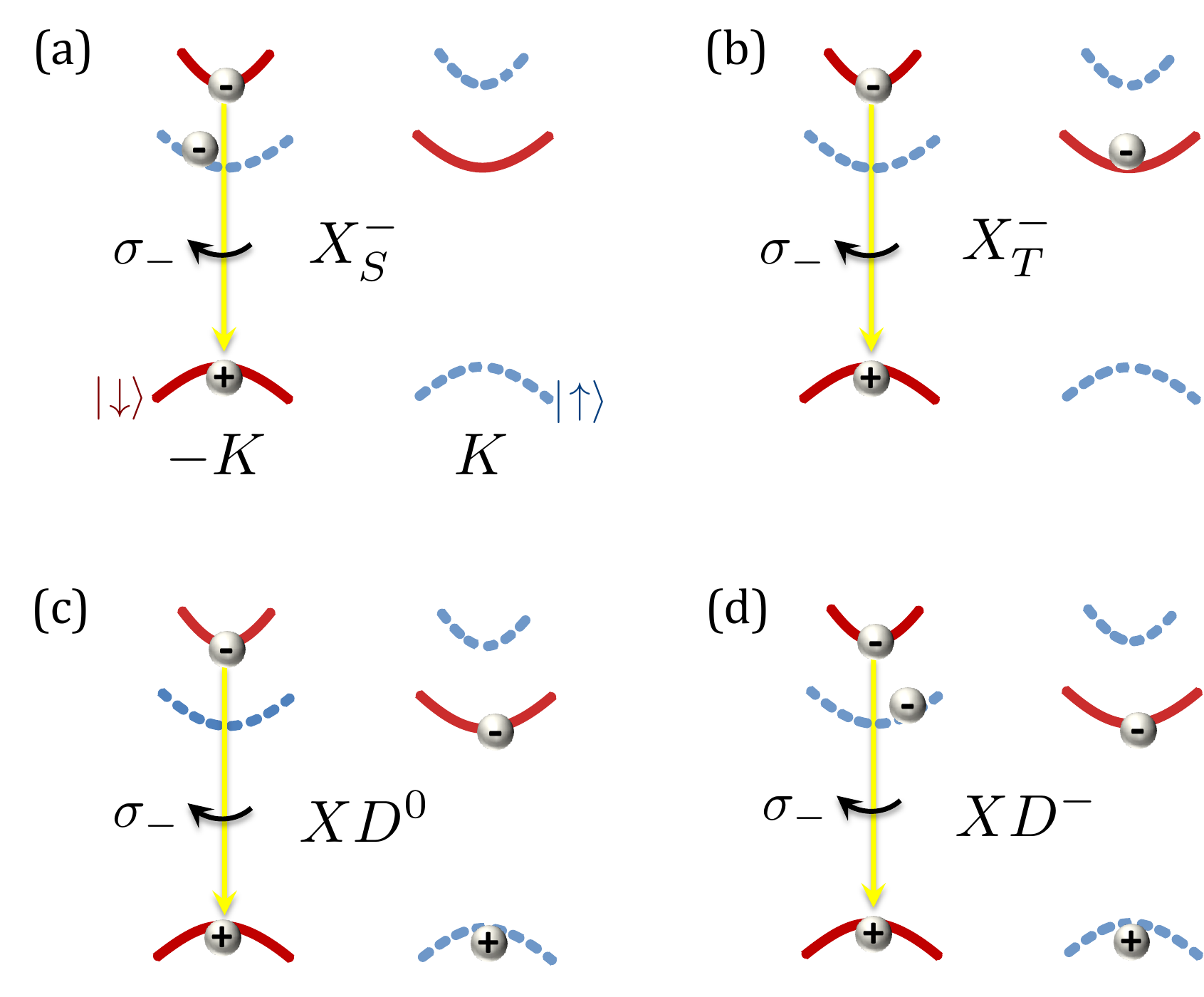}
\caption{Trions and biexcitons with bright exciton component in the $-K$ valley. (a)/(b) Negative bright trions with singlet/triplet configuration of their electrons. (c)/(d) Neutral and charged biexcitons.
\label{fig:345}}
\end{figure}

The other resonances seen in Fig.~\ref{fig:exp} are singlet and triplet bright trions, as well as neutral and charged biexcitons. Their compositions are depicted in Fig.~\ref{fig:345}. the singlet or intravalley trion, $X_S^-$,  involves electrons from the same valley but opposite spin, whereas the triplet or intervalley trion, $X_T^-$,  involves electrons from opposite valleys but similar spin.  The formation of negative trions imply that the monolayer has a finite density of free electrons, which are either residual at zero-gate voltage or that are introduced by the photoexcitation. The latter is feasible given that the laser energy, 1.96~eV, is above the band-gap energy of the free electron-hole continuum at 1.9~eV in monolayer WSe$_2$ that is encapsulated by hexagonal boron nitride \cite{Stier_PRL18}. Figure~\ref{fig:exp} shows a strong co-polarized emission from the triplet trion under circularly-polarized photoexcitation. The reason for this behavior is the valley pumping effect \cite{Robert_NatComm21}. 

The biexciton and charge biexciton resonances in Fig.~\ref{fig:exp} emerge at large laser powers.  The emission intensity of the biexciton resonance, $XD^0$,  has a quadratic dependence on laser power, whereas the charged biexciton, $XD^-$, has a sub-quadratic dependence  \cite{Chen_NatComm18,Ye_NatComm18,Li_NatComm18,Barbone_NatComm18}. The quadratic dependence of the neutral biexciton is clear since it takes two excitons to form a biexciton. The origin for the sub-quadratic dependence of the charge biexciton is more subtle, and depends on the interplay between the densities of free electrons ($n_e$), photoexcited electron-hole pairs ($n_x$) and trions ($n_t$). For a given pair density, the formation rate of trions from excitons is optimal when $n_e \gg n_x$ whereas it is hampered when $n_e \sim n_x$. The latter condition is applicable when the laser power is large and the monolayer is nearly charge neutral. The slowdown in trion formation when $n_e \sim n_x$ leads to a slowdown in formation of charged biexcitons ($\propto\,n_xn_t$),  rendering their sub-quadratic dependence on laser power. On the other hand, when $n_e \gg n_x, n_t$, the photoexcited pair is more likely to become a trion rather than a charged biexciton ($ n_xn_e \gg n_xn_t$). This behavior explains why the emission intensity of $XD^-$ decays fast when electrons are added to the monolayer \cite{Chen_NatComm18,Ye_NatComm18,Li_NatComm18,Barbone_NatComm18}, whereas the emission from trions saturates.  

\section{First principle calculation of the band structure and eigenstates}
\label{appendix:DFT}
The Bloch wavefunctions are calculated by using the \textit{ab initio} {\sc Quantum ESPRESSO} package \cite{QE}, in which the norm-conserving full relativistic pseudopotentials with the PBEsol exchange correlation are obtained from \url{www.pseudo-dojo.org} \cite{PseudoDojo}. The lattice constant is $a_0$$\,=\,$3.33$\,\AA$ (3.19$\,\AA$) for WSe$_2$ (WS$_2$), and the distance between the chalcogen sublayers is $d_{x-x}\,=\,$3.36$\,\AA$ (3.15$\,\AA$). The distance between van der Waals layers is set at 5.18$a_0$ to suppress interlayer coupling. The  energy  cutoffs  for  wavefunction  and  charge density are 40 and 180~Ry, respectively.

%
%
%
\section{Character table of the $C_{3h}$ double group}
\label{appendix:B}
\begin{table} [h]
\numberwithin{table}{section}
\caption{Character table of the $C_{3h}$ double point group. The electronic states at $\pm K$ transform like the IRs $K_7$-$K_{12}$, as shown in Fig.~\ref{fig:recom}(a). In-plane vector components transform as $K_2$ and $K_3$, the out-of-plane vector component as $K_4$, and zone-edge phonon modes that couple same-spin states in the conduction (valence) band transform as $K_3$ ($K_1$). Here $\omega = e^{2\pi i/3}$.}
\label{tab:B1}
\renewcommand{\arraystretch}{1.5}
\begin{tabular}{c|cccccccccccc|c}
\hline 
$C_{3h}$ & $E$ & $C_3^+$ & $C_3^-$ & $\sigma_h$ & $S_3^+$  & $S_3^-$ & $\bar{E}$ & $\bar{C}_3^+$ & $\bar{C}_3^-$ & $\bar{\sigma}_h$ & $\bar{S}_3^+$  & $\bar{S}_3^-$ &   \\ \hline \hline
$K_1$ & 1 & 1 & 1 & 1 & 1 & 1 & 1 & 1 & 1 & 1 & 1 & 1 & 1 \\
$K_2$ & 1 & $\omega$ & $\omega^*$ & 1 & $\omega$ & $\omega^*$ & 1 & $\omega$ & $\omega^*$ & 1 & $\omega$ & $\omega^*$ & \\
$K_3$ & 1 & $\omega^*$ & $\omega$ & 1 & $\omega^*$ & $\omega$ & 1 & $\omega^*$ & $\omega$ & 1 & $\omega^*$ & $\omega$ & \\
$K_4$ & 1 & 1 & 1 & -1 & -1 & -1 & 1 & 1 & 1 & -1 & -1 & -1 & \\
$K_5$ & 1 & $\omega$ & $\omega^*$ & -1 & -$\omega$ & -$\omega^*$ & 1 & $\omega$ & $\omega^*$ & -1 & -$\omega$ & -$\omega^*$ & \\
$K_6$ & 1 & $\omega^*$ & $\omega$ & -1 & -$\omega^*$ & -$\omega$ & 1 & $\omega^*$ & $\omega$ & -1 & -$\omega^*$ & -$\omega$ & \\
$K_7$ & 1 & -$\omega$ & -$\omega^*$ & $i$ & -$i\omega$ & $i\omega^*$ & -1 & $\omega$ & $\omega^*$ & -$i$ & $i\omega$ & -$i\omega^*$ & \\
$K_8$ & 1 & -$\omega^*$ & -$\omega$ & -$i$ & $i\omega^*$ & -$i\omega$ & -1 & $\omega^*$ & $\omega$ & $i$ & -$i\omega^*$ & $i\omega$ & \\
$K_9$ & 1 & -$\omega$ & -$\omega^*$ & -$i$ & $i\omega$ & -$i\omega^*$ & -1 & $\omega$ & $\omega^*$ & $i$ & -$i\omega$ & $i\omega^*$ & \\
$K_{10}$ & 1 & -$\omega^*$ & -$\omega$ & $i$ & -$i\omega^*$ & $i\omega$ & -1 & $\omega^*$ & $\omega$ & -$i$ & $i\omega^*$ & -$i\omega$ & \\
$K_{11}$ & 1 & -1 & -1 & $i$ & -$i$ & $i$ & -1 & 1 & 1 & -$i$ & $i$ & -$i$ & \\
$K_{12}$ & 1 & -1 & -1 & -$i$ & $i$ & -$i$ & -1 & 1 & 1 & $i$ & -$i$ & $i$ & \\ \hline\hline
\end{tabular}
\end{table}

%
%
%
%
%


\begin{thebibliography}{99}
\bibitem{comment_pol} In principle, the photoexcited bright exciton can exhibit valley coherence, and thus linearly polarized emission following linearly-polarized photoexcitation. The reason for similar emission amplitudes of the bright exciton with co-polarized and cross-polarized light in our experiment is the loss of valley coherence. Valley decoherence is especially relevant when the photoexcitation generates free electron hole pairs in the continuum, which is the case when WSe$_2$ is photoexcited by HeNe laser (1.96~eV) if the monolayer is encapsulated by hexagonal boron nitride. The energy gap for free electron hole pairs of this system is $\sim$1.9 eV  \cite{Stier_PRL18}. %
\end{thebibliography}
\end{document}